\documentclass[12pt,scrartcl,a4paper]{article}
\usepackage{a4wide}
\usepackage{setspace}
\usepackage{epsfig}

\usepackage[affil-it]{authblk}

\usepackage{amsfonts,amssymb,graphics,epsfig,epstopdf,verbatim,bm,latexsym,amsmath,url,amsbsy,paralist,natbib,xcolor}

\numberwithin{equation}{section}

\newcommand{\Cov}{{\rm Cov}}

\newcommand{\R}{\mathbb{R}}
\newcommand{\N}{\mathbb{N}}

\newcommand{\E}{\mathrm {E}}

\newcommand{\CvM}{{\rm \operatorname {CvM}}}
\newcommand{\I}{{\rm \operatorname I}}

\newtheorem{theorem}{Theorem}
\newtheorem{lemma}{Lemma}

\newcounter{bean}

\newcommand\blfootnote[1]{%
  \begingroup
  \renewcommand\thefootnote{}\footnote{#1}%
  \addtocounter{footnote}{-1}%
  \endgroup
}

\begin{document}

\title{Testing marginal homogeneity in Hilbert spaces with applications to stock market returns}

\author{Marc Ditzhaus$^{\dagger}$ and
	Daniel Gaigall$^{\ddagger}$}

\affil{$^{\dagger}$ Department of Statistics, TU Dortmund University, 44221 Dortmund, Germany.\\$^{\ddagger}$ Institute of Actuarial and Financial Mathematics, Leibniz University Hannover, 30167 Hannover, Germany.}

\date{}

\maketitle

\blfootnote{Email adresses: $^{\dagger}$marc.ditzhaus@tu-dortmund.de, $^{\ddagger}$gaigall@stochastik.uni-hannover.de}

\begin{abstract}
\noindent

The paper considers a  paired data framework  and discuss the question of marginal homogeneity of  bivariate high dimensional or functional data.  The related testing problem can be endowed into a more general  setting for paired random variables taking values in a general Hilbert space. To address this problem,  a Cram\'er-von-Mises type test statistic is applied and  a bootstrap   procedure is suggested to obtain critical values and finally a  consistent test. The desired properties of a bootstrap test can be derived, that are asymptotic exactness under the null hypothesis and consistency under  alternatives. Simulations show the quality of the test in the finite sample case. A possible application is  the comparison of two possibly dependent  stock   market returns on the basis of functional data. The approach is demonstrated  on the basis of  historical data for different stock market indices.


\end{abstract}

	\section{Introduction}

Due to the availability of high frequency data, statistical observations can be described and modeled by random functions, so-called stochastic processes. Since classical methods are designed for vector-valued observations rather than for stochastic processes, they usually cannot be applied in this situation. The field of functional data tries to close this gap. One popular solution to tackle this problem is to project the random functions to the real line and then apply one of the classical methods. For example, \cite{cue2006} and \cite{cue2007} applied the Kolomogorov--Smirnov goodness-of-fit test to randomly projected square integrable functions. \cite{cuevas} extended this idea to more general spaces. \cite{DitzhausGaigall2018} did the same by discussing observations with values in a general Hilbert space. In this way, stochastic processes as well as high-dimensional data can be discussed simultaneously. More interaction between these two fields is desirable as stated by \cite{goiaVieu2016} and \cite{cuevas14} in the functional data community as well as by \cite{ahmed2017} from the high-dimensional side. Extending the idea of goodness-of-fit, \cite{bun} studied the testing problem whether the underlying distribution belongs to a pre-specified parametric family. \cite{hal2017} discussed pre-processing the functional data, which is in practice usually just available at finitely many time points, in the context of two-sample testing. Recently, the two-sample testing problem for multivariate functional data was discussed by 	\cite{jiang2017} on the basis of empirical characteristic functions. 

\medskip
In this paper, we use also the projection idea mentioned but we address a paired-sample testing problem allowing for dependency. This approach and our test developed are  completely new. We suggest a procedure for testing marginal homogeneity in separable Hilbert spaces, where we consider not just a few random projects but all projects from a sufficient large projection space.   The advantage of our approach is that no additional randomness has an influence on the result of the test. With a view to the consistency of the testing procedure, we apply a  test statistic  of  Cram\'er-von-Mises type.  See \cite{and} and  \cite{ros} for Cram\'er-von-Mises tests in the usual cases of real-valued random variables and random vectors with real components, and  \cite{gai} for the application of a Cram\'er-von-Mises test for the null hypothesis of marginal homogeneity for bivariate distributions on the Cartesian square of the real line. However, the demand for consistency has a price:  the distribution of the test statistic under the null hypothesis is unknown and so related quantiles are not available in practice. In contrast to the unpaired setting, exchangeability is not given in general under the null hypothesis of marginal homogeneity. This  is already known for bivariate distributions on the Cartesian square of the real line, see \cite{gai}. For that reason, a permutation test, such as they are  used by \cite{hal} and \cite{BugniHorwitz2018} for the unpaired two-sample setting under functional data,  or in  \cite{BugniHorwitz2018} in the more general situation of one control group against several treatment groups, is no option in our situation. To solve this problem, we offer a bootstrap   procedure to determine critical values. We can show that the the bootstrap  test keeps the nominal level asymptotically under the null hypothesis. Moreover, we prove the consistency of our approach with respect to the bootstrap  procedure under any alternative. 

\medskip
While also other applications are possible, for instance to high dimensional data, we especially focus on functional data and stock market returns. In the field of empirical finance, statistical inference of  stock market prices is a widespread topic. Here, usual statistical procedures are typically applied to stock market log returns. One popular question is that of a suitable distributional model which matched with the observations. For example,  \cite{goe} apply different goodness-of-fit tests to data sets consist of daily stock market index returns for several emerging and developed markets and \cite{gra} consider the distribution of log stock index returns of the S\&P 500 and deduce that the distributions do not follow a normal distribution but demonstrate a greater ability for other distributional models. This topic is also treated on the basis of high frequency data, see \cite{mal}, where 5 min  returns of the Nasdaq Composite index and 1 min returns of the S\&P 500 are considered. Besides the topic of model selection for stock market returns, another interesting topic is the comparison of different stock market returns such as it is done in \cite{mid}, where annual pooled data of 100 conventional and Islamic stock returns are analyzed. In this context, dependencies between different stock prices are obvious and already detected, see \citet{min}, where the major 8 companies of the Korean stock market are investigated, and  should be taken into account.

\medskip
The paper is structured as follows. We first introduce the model and our general null hypothesis of marginal homogeneity in the paired sample setting in Section \ref{sec:model}. 
We introduce a Cram\'er-von-Mises type test for the aforementioned testing problem and derive its asymptotic behavior with the help of the theory of $U$-statistics. The resulting asymptotic law under the null hypothesis can be transfered to a bootstrap counterpart of the test statistic. In addition to these theoretical findings, we study the small sample performance of the two resampling tests in a numerical simulation study presented in Section \ref{sec:sim}. Finally, the application to stock market indices is outlined in Section \ref{sec:data_exam}. We demonstrate the application of the test to the  historical values of the stock market indices Nikkei Stock Average from Japan, Dow Jones Industrial Average from US, and Standard \& Poor’s 500 from US. The test confirm the intuition that the indices of the same county are much more comparable than indices of different countries. Note that all proofs are conducted in the Appendix.

    \section{Testing marginal homogeneity in Hilbert spaces }\label{sec:model} 
    Let $H$ be a  Hilbert space, i.e. a real inner product space, where the inner product is denoted by $\langle \cdot,\cdot \rangle$. We suppose that $H$ is separable with    countable orthonormal basis $O=\lbrace e_i;i\in I\rbrace$, where $e_i$ is the $i$-th basis element and the index set $I$ is given by the natural numbers $I=\N$ or the subset $I=\{1,\dots,|I|\}\subset\N$.  Now, let paired observations be given
    \begin{align*}
    	X_{j}= (X_{j,1},X_{j,2} ),~j=1,\dots,n,
    \end{align*}
 that are random variables with values in  $H\times H$.  We suppose that $X_1,\ldots,X_n$ are independent and identical distributed, and we suppose that the distribution   $P^{X_1}$ of $X_1$ is unknown.  For technical reasons, we suppose that  for all $i\in I$ the joint distribution of $\langle X_{1,1},e_i\rangle$ and $\langle X_{1,2},e_i\rangle$ is absolutely continuous with density $f_i$, where the set $\{f_i(r,s)>0;(r,s)\in\R^2\}$ is open and convex, compare with \cite{gai}. While we allow any dependence structure between $X_{j,1}$ and $X_{j,2}$, we like to infer the null hypothesis of marginal homogeneity
    \begin{align*}
    	\mathcal H: P^{X_{j,1}} =  P^{X_{j,2}} \quad \text{versus}\quad \mathcal K: P^{X_{j,1}} \neq  P^{X_{j,2}}.
    \end{align*}
	As postulated in the introduction, we project first the processes $X_{j,i}$ to the real line and then apply a Cram\'er-von-Mises type test. Projection is done via the inner product, i.e., we consider $\langle X_{j,i},x\rangle$ for $x\in H$. We consider all projections $x$ from a sufficient large projection space $h\subset H$. In fact, as explained in \cite{DitzhausGaigall2018}, the distributions of $X_{1,1}$ and $X_{1,2}$ coincide if and only if $\langle X_{1,1}, x\rangle$ and $\langle X_{1,2}, x \rangle$ have the same distribution for all projections $x\in h$, where
	\begin{align*}
		h = \Bigl\{ \sum_{j=1}^k m_{j} e_{i_j}; k\in I,i_1,\ldots, i_k\in I, i_1< \ldots < i_k, \sum_{j=1}^k m_{j}^2 = 1  \Bigr\}.
	\end{align*}
	This motivates the following test statistic:
	\begin{equation}\label{eqn:CVM}
		\CvM_n=\int D_n(x)\mathcal P(\mathrm d   x),
	\end{equation}
	where $\mathcal P$ is a suitable probability measure on the projection space $h$ and $D_n(x)$ is the usual two-sample Cram\'er-von-Mises  distance when applying the projection $x\in h$. Let 
	\begin{equation*}
	\begin{split}
	F_{n,i }(x,r)=\frac 1 n\sum_{j=1}^n \I_{\langle x,X_{j,i}\rangle \le r},~(x,r)\in H\times \R, ~i=1,2,
	\end{split}
	\end{equation*}
	be the empirical distribution function of the real-valued random variables \linebreak $ \langle x,X_{1,i}\rangle,\dots,\langle x,X_{n,i}\rangle$. Then the related  two-sample  Cram\'er-von-Mises distance is given by
	\begin{equation*}
	\begin{split}
	D_n(x)= n  \int [ F_{n,1}(x,r)- F_{n,2}(x,r)]^2\bar F_n(x,\mathrm d r),
	\end{split}
	\end{equation*}
	where $\bar F_n = (F_{n,1}+ F_{n,2})/2$.
	The probability measure $\mathcal P$ can be chosen arbitrarily in advance as long as some regularity assumptions are fulfilled. While more general measures $\mathcal P$ may be considered, we focus here to the following specific proposal. It is based on two probability measures $\nu_1$ and $\nu_2$ on the index set $I$ such that $\nu_j(\{i\})>0$  for all $i\in I$. In the case of functional data, we can choose Poisson distributions shifted by 1, for instance. In what follows, we specify  the probability measure $\mathcal P$ by determining the  procedure to generate a  realization of $\mathcal P$. This procedure is also useful  to obtain the concrete value of the test statistic by Monte-Carlo simulation in applications.  
	\setcounter{bean}{0}
	\begin{center}
		\begin{list}
			{\textsc{Step} \arabic{bean}.}{\usecounter{bean}}
			\item Generate a realization $k\in I$ of the distribution $\nu_1$.
			\item Independently of Step 1, generate $i_1,\dots,i_k\in I$ by $k$-times sampling without replacement from the distribution $\nu_2$.
			\item Independently of Steps 1 and 2, generate a realization $(m_1,\dots,m_k)$ of the uniform distribution on the unit circle in $\R^{k}$.
			\item Set $x=\sum_{j=1}^k m_je_{i_j}$.
		\end{list}
	\end{center}

	\subsection{Asymptotic theory of the test statistic}\label{sec:asy_uncon}
	For our asymptotic approach, we let $n\to\infty$. It is well known that the Cram\'er-von-Mises distance $D_n$ is connected to von Mises' type functionals, also known as $V$-Statistics, which are closely related to $U$-Statistics. For a deeper introduction to these kinds of statistics, we refer the reader to \cite{kor} and \cite{serf}. Our statistic $\CvM_n$ can also be rewritten into a certain $V$-Statistic and, thus, the same theory can be applied to obtain the following result.	
	\begin{theorem}\label{theo:asym_S}
		Let $\tau_1,\tau_2,\ldots$ be a sequence of independent standard normal distributed random variables. Under the null hypothesis $\mathcal H$,
		\begin{align}\label{eqn:null_conv}
			\CvM_n \overset{ d}{\to} \sum_{i=1}^\infty \lambda_i (1+\tau_i^2) = Z,
		\end{align}
		where $(\lambda_i)_{i\in\N}$ is a sequence of non-negative numbers with $\sum_{i=1}^\infty \lambda_i< \infty$ and $\lambda_{i}>0$ for at least one $i\in\N$ implying that  the distribution function of  $Z$ is continuous and strictly increasing on the non-negative half-line.
	\end{theorem}
	\begin{theorem}\label{theo:cons_S}
		Under the alternative $\mathcal K$, our statistic  $\CvM_n$ diverges, i.e., $\CvM_n\overset{p}{\rightarrow}\infty$ as $n\to\infty$. 
	\end{theorem}
In general, the test statistic $\CvM_n $ is not distribution-free under the null hypothesis, i.e., the distribution depends on the unknown distribution of $X_1$. As it can be seen in the proofs, the same applies to    $Z$. Given that  $\alpha\in(0,1)$ is the significance level, neither a  $(1-\alpha)$-quantile $c_{n,1-\alpha}$ of $\CvM_n $ nor the $(1-\alpha)$-quantile $c_{1-\alpha}$  of  $Z$  is available as critical value in applications. To resolve this problem, we propose the estimation of the quantiles via bootstrapping in the spirit of \cite{efron} and  follow the idea in \cite{gai}, where the usual two-sample Cram\'er-von-Mises distance is applied to  bivariate random vectors with values in  $\R^2$.  Note that under the null hypothesis $\mathcal H$ the expectations $\E[ F_{n,1}(x,y)]=\E[ F_{n,2}(x,y)],~(x,y)\in H\times \R$, coincide and, thus, we can rewrite our test statistic into
	\begin{equation*}
		\CvM_n=n\int  \int \{ F_{n,1}(x,y)-\E[ F_{n,1}(x,y)]+\E[ F_{n,2}(x,y)]- F_{n,2}(x,y)\}^2\bar F_n(x,\mathrm d y)\mathcal P(\mathrm d   x).
	\end{equation*}
	Denote by $X_{jn}^*= (X_{jn,1}^*,X_{jn,2}^* ),~j=1,\dots,n$, a bootstrap sample from the original observations $X_j$, $j=1,\dots,n$, obtained by $n$-times sampling with replacement. Let $F_{n,i}^*$, $\bar F_{n}^*$ be the bootstrap counterparts of $F_{n,i}$ and $\bar F_n$. Clearly, the conditional expectation $\E[ F_{n,i}^*(x,y)\vert(X_j)_j]$ given the data $(X_j)_{j=1,\ldots,n}$ equals $F_{n,i}(x,y)$. Consequently, the bootstrap counterpart of our test statistic is
	\begin{equation*}
	\CvM_{n}^*=n\int  \int\Big( F_{n,1}^*(x,y)- F_{n,1}(x,y)+ F_{n,2}(x,y)- F_{n,2}^*(x,y)\Big)^2\bar F_{n}^*(x,\mathrm d y)\mathcal P(\mathrm d   x).
	\end{equation*}
	Let $c_{n,1-\alpha}^*$ be a $(1-\alpha)$-quantile of $\CvM_n^*$ given the original observations $X_1,\ldots,X_n$. In applications, concrete values of $c_{n,1-\alpha}^*$ are obtained by Monte-Carlo simulation. In the proofs, we  show that the bootstrap statistic mimics asymptotically the limiting null distribution under the null hypothesis $\mathcal H$ implying that $c_{n,1-\alpha}^*$ is an appropriate estimator for the unknown quantile  $c_{n,1-\alpha}$ or $c_{1-\alpha}$, while $\CvM_n^*$ and $c_{n,1-\alpha}^*$ remain asymptotically finite under general alternatives. This results in an asymptotically exact and consistent bootstrap test $\varphi_n^*=\I_{\CvM_n>c_{n,1-\alpha}^*}$.	
	\begin{theorem}\label{theo:boot}
		As $n \to \infty$ we have 	$\E[\varphi_n^*]=P(\CvM_n>c_{n,1-\alpha}^*) \to \alpha \I_{\mathcal H}+  \I_{\mathcal K}$, where $\I_{\cdot}$  denotes the indicator function.
	\end{theorem}

%
%
%
%
%

	\section{Simulations}\label{sec:sim}
	\setcounter{equation}{0}
	
	\renewcommand{\arraystretch}{1}
	\begin{table}[tb]
		\caption{\label{tab1}Empirical sizes.}
		\begin{center}
			\begin{tabular}{@{}ccccc@{}}
				$X_{j,1}(t)$&$X_{j,2}(t)$&$r$ &   $\alpha = 5\%$  & $\alpha = 10\%$ \\
				\hline

				$B_{j,1}(t)$ &$B_{j,2}(t)$ &  0  & 5.2 &  10.7 \\
				& &  0.25 &   3.4 &  9.0 \\
				& &  0.5 &  3.9 &    8.4 \\
				$1.5 B_{j,1}(t)$ &$1.5 B_{j,2}(t)$  & 0  & 4.5 & 9.6  \\
				& &  0.25 &   4.4 & 9.1 \\
				& &  0.5 & 4.0  &   7.9 \\
				$2B_{j,1}(t)$ &$2B_{j,2}(t)$ & 0  & 5.2 &  12.0 \\
				& &  0.25  & 3.9 &  8.5 \\
				& &  0.5  & 3.7 & 7.1 \\
				$2.5B_{j,1}(t)$ &$2.5B_{j,2}(t)$ & 0 & 5.5 &10.3 \\
				& &  0.25  & 5.5  & 10.3\\
				& &  0.5 &    4.3 & 7.6 \\
				$B_{j,1}(t)+0.5 t (1-t)$ &$B_{j,2}(t)+0.5 t (1-t)$ & 0 &  5.1 & 10.0 \\
				& &  0.25 &  4.7 & 9.9  \\
				& &  0.5 & 3.2 &  7.2  \\
				$B_{j,1}(t)+t (1-t)$ &$B_{j,2}(t)+ t (1-t)$ & 0   & 5.2 & 10.3   \\
				& &  0.25 &   4.0 &  8.3 \\
				& &  0.5 & 2.6 &  6.4 \\
				$B_{j,1}(t)+1.5 t (1-t)$ &$B_{j,2}(t)+1.5 t (1-t)$ & 0  &   5.2 &  9.9  \\
				& &  0.25 &   5.1 &   9.5\\
				& &  0.5 &  2.9 &  7.7\\
				$B_{j,1}(t)+2 t (1-t)$ &$B_{j,2}(t)+2t (1-t)$ & 0 & 5.8& 10.7  \\
				& &  0.25 & 5.2 &  11.2\\
				& &  0.5 &  3.7 &  8.1 
			\end{tabular}
		\end{center}
	\end{table}

\begin{figure}
		\begin{minipage}{.48\textwidth}
			\begin{flushright}
			{ \footnotesize $X_{j,1} (t)= B_{j,1}(t)$,  $X_{j,2} (t)= 1.5 B_{j,2}(t)$ \quad}
			\includegraphics[width=1.0\linewidth]{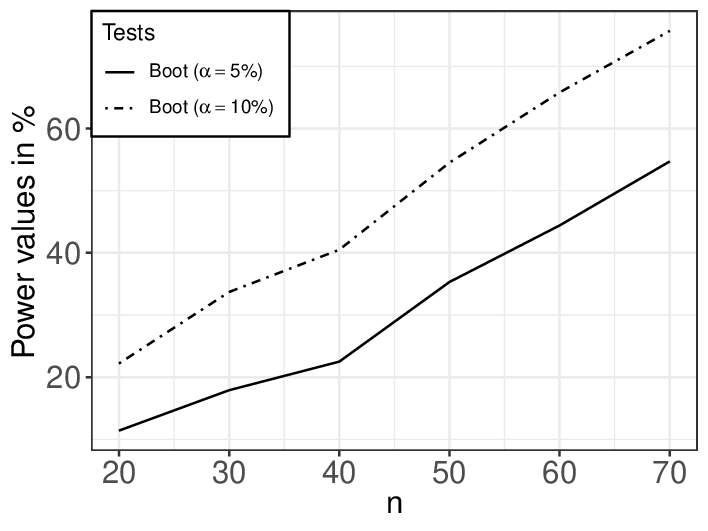}
		\end{flushright}
		\end{minipage}%
		\quad
		\begin{minipage}{.48\textwidth}
			\begin{flushright}
				{ \footnotesize $X_{j,1} (t)= B_{j,1}(t)$, $X_{j,2}(t) = B_{j,2}(t) +t(1-t)$}
			
			\includegraphics[width=1.0\linewidth]{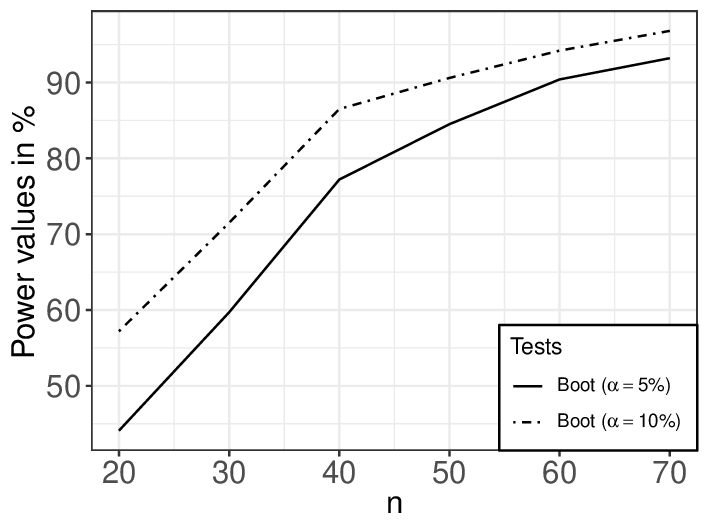}
		\end{flushright}
		\end{minipage}
		\caption{{Power values for increasing sample sizes under moderate ($r=0.25$) dependency. }}
		\label{fig:incrn}
	\end{figure}

	\renewcommand{\arraystretch}{1}
	\begin{table}[tb]
		\caption{\label{tab2} Empirical power values.}
		\begin{center}
			\begin{tabular}{@{}ccccc@{}}
				$X_{j,1}(t)$&$X_{j,2}(t)$&$r$&  $\alpha = 5\%$ & $\alpha = 10\%$ \\
				\hline
				$ B_{j,1}(t)$ &$1.5 B_{j,2}(t)$ & 0 & 10.7 & 19.9 \\
				& &  0.25 & 10.0 & 21.0 \\  
				& &  0.5 &  12.0 & 25.2\\
				$B_{j,1}(t)$ &$2B_{j,2}(t)$ & 0 & 22.6 & 42.9   \\ 
				& &  0.25 &  36.3 & 58.7\\
				& &  0.5 & 50.6 &  73.1\\
				$B_{j,1}(t)$ &$2.5B_{j,2}(t)$ & 0 & 47.2 &  74.3  \\
				& &  0.25 &  62.5 & 85 \\
				& &  0.5 &  84.7  & 94.3\\
				$B_{j,1}(t)$ &$B_{j,2}(t)+0.5 t (1-t)$ & 0 & 14.1 & 21.2 \\
				& &  0.25 & 13.0 & 21.6\\
				& &  0.5 & 17.0 &  24.4 \\
				$B_{j,1}(t)$ &$B_{j,2}(t)+ t (1-t)$ & 0 & 35.2 &46.9\\
				& &  0.25 &  45.1 & 57.0\\ 
				& &  0.5 & 52.8 & 66.4\\
				$B_{j,1}(t)$ &$B_{j,2}(t)+1.5 t (1-t)$ & 0 & 68.7 &  78.8  \\
				& &  0.25 &  78.7 & 86.4\\
				& &  0.5 &  87.8 & 93.5\\
				$B_{j,1}(t)$ &$B_{j,2}(t)+2t (1-t)$ & 0 &  92.6 &96.0 \\
				& &  0.25 &  95.8 & 97.4\\
				& &  0.5 &  98.9 &  99.4
			\end{tabular}
		\end{center}
	\end{table}
	 Remembering that our test is suitable for random variables  $X_{i,j}$, $i=1,2$, $j=1,\dots,n$, with values in a general separate Hilbert space, we consider the separable Hilbert space $ H$ consisting of all measurable and square integrable functions on the unit interval $[0,1]$. This space is endowed with the usual inner product $\langle \cdot,\cdot \rangle$ and the normalized Legendre polynomials build a corresponding orthonormal basis $O=\lbrace e_i;i\in I\rbrace$, $I=\N$. 
	We obtain  our test statistic \eqref{eqn:CVM} by Monte-Carlo simulation based on  500 replications following {\textsc{Step}} 1--4 from Section \ref{sec:model}. Thereby, we choose in {\textsc{Step}} 1 and {\textsc{Step}} 2 a standard Poisson distribution shifted by $1$, i.e., the distribution of $N+1$ for $N\sim \text{Pois}(1)$. In our simulations, the stochastic processes  $X_{j,i}=(X_{j,i}(t);t\in [0,1])$ have the form
	\begin{align*}
	X_{j,i}(t)=a_iB_{j,i}(t)+b_it(t-1),~t\in [0,1],~i=1,2,~j=1,\dots,n
	\end{align*}
	for parameters  $a_i\in \R\setminus \{0\}$ and $b_i\in\R$  and independent bivariate Brownian bridges $B_j=(B_{j,1}, B_{j,2})$ on $[0,1]$, $j=1,\dots,n$, with covariance structure 
	\begin{align*}
	\Cov(B_{j,1}(s),B_{j,2}(t))=r(\min(s,t)-st),~s,t\in[0,1],~j=1,\dots,n
	\end{align*}
	for a dependency parameter $r\in[0,1]$.  Each simulation is based on 1000 simulation runs. To obtain the critical values in the bootstrap procedure, we use  Monte-Carlo simulation based on  999  replications. Empirical size and power values of the bootstrap  test are displayed in Tables \ref{tab1} and \ref{tab2}, respectively. The simulations are conducted for parameters $r\in\{0,0.25,0.5\}$, $a_i\in\{1,1.5,2,2.5\}$, and $b_i\in\{0,0.5,1,1.5,2\}$, the sample size $n=20$, and  significance levels $\alpha\in\{5 \%, 10 \%\}$. The  empirical sizes are in almost all cases in a reasonable range around the nominal level $\alpha$. A systematic exception from this observations are the sizes of the bootstrap approach under the strong dependence setting $(r=0.5)$. In this case, the bootstrap decisions are rather conservative with corresponding empirical sizes from $2.6\%$ to $4.3\%$ with an average of $3.5\%$ for $\alpha = 5\%$ as well as values from $6.4\%$ up to $8.4\%$ and an average of $7.6\%$ for $\alpha = 10\%$. Regarding the  data example discussed in the upcoming section, we primarily studied here the sample size setting $n=20$.  To show that the power values grow for increasing sample sizes $n\in\{20,30,\ldots,70\}$, we conducted additional simulations for two specific alternatives $X_{j,1} (t)= B_{j,1}(t)$ and  $X_{j,2} (t)= 1.5 B_{j,2}(t)$ as well as $X_{j,1} (t)= B_{j,1}$ and $X_{j,2}(t) = B_{j,2}(t) +t(1-t)$, $t\in[0,1]$, $j=1,\dots,n$, under moderate $(r=0.25)$ dependency, see Figure \ref{fig:incrn} for the results.

	\section{Applications to stock market returns} \label{sec:data_exam}

\begin{figure}
   	\begin{minipage}{.5\textwidth}
   		\centering
   		\includegraphics[width=1.0\linewidth]{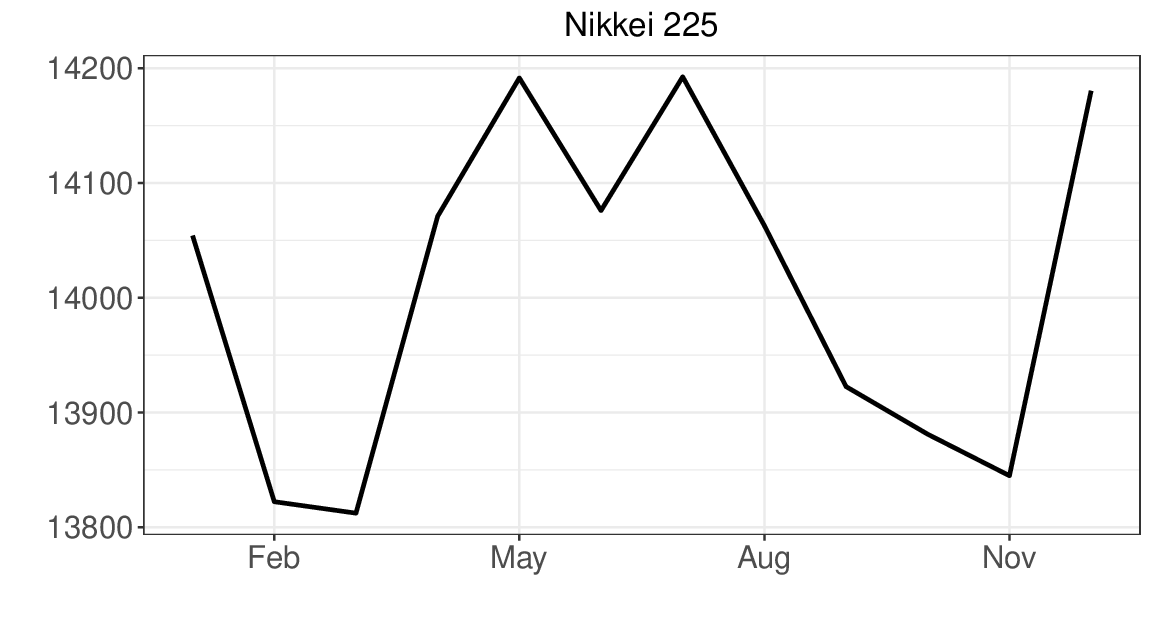}
   	\end{minipage}%
   	\begin{minipage}{.5\textwidth}
   		\centering
   		\includegraphics[width=1.0\linewidth]{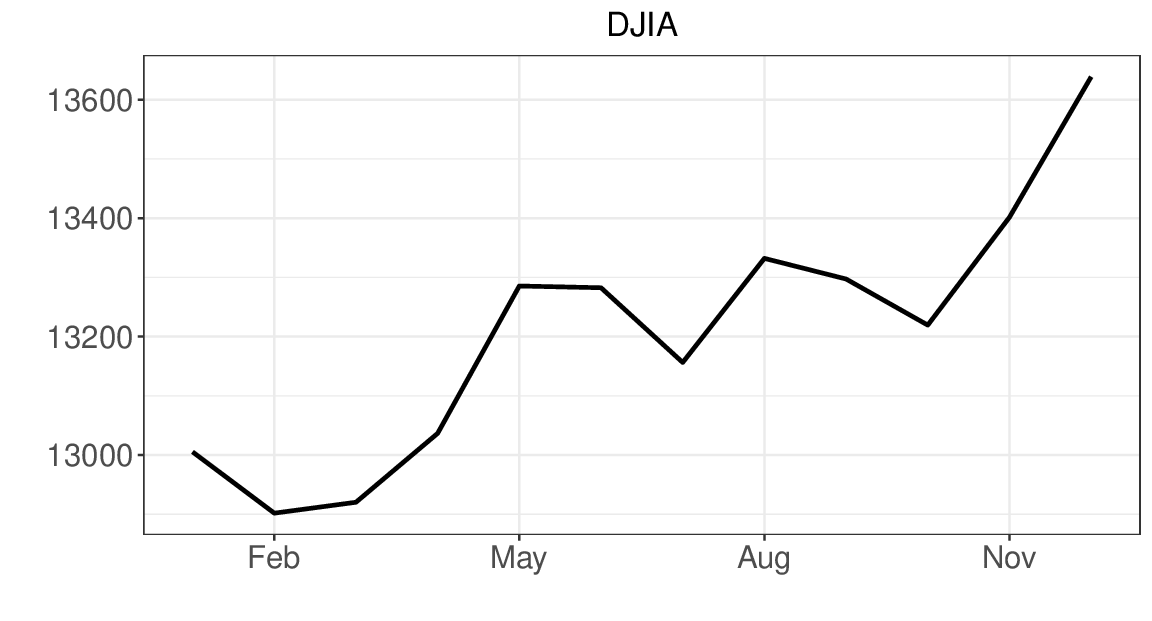}
   	\end{minipage}
   \centering
   	\begin{minipage}{.5\textwidth}
   		\centering
   		\includegraphics[width=1.0\linewidth]{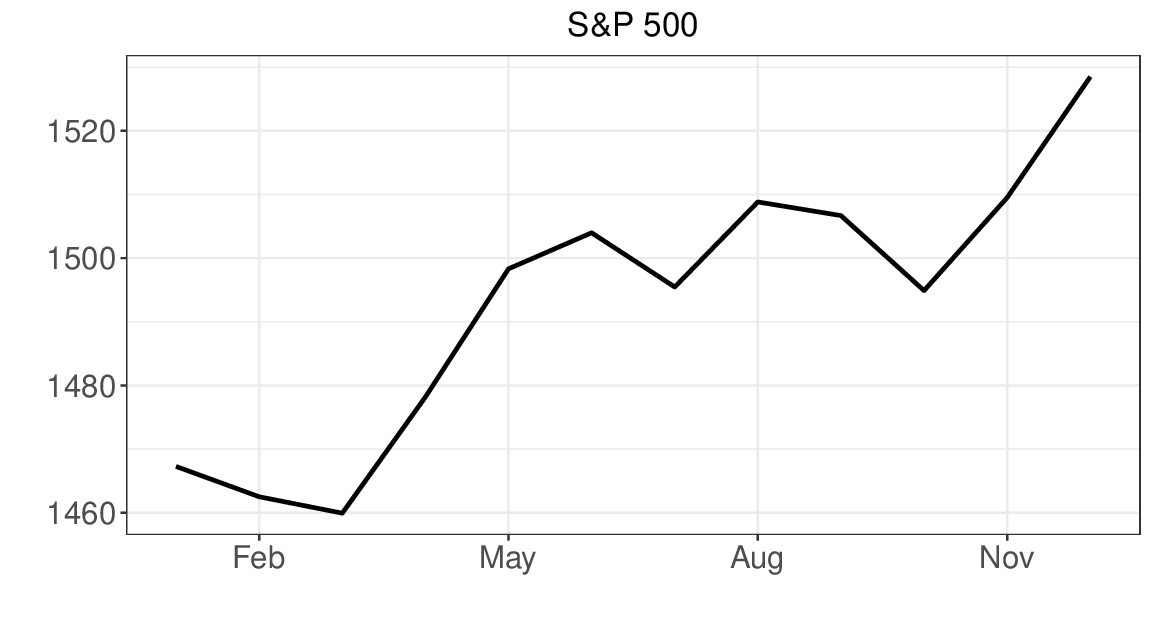}
   	\end{minipage}%
   	\caption{{Mean monthly values from 01/01/1999 to 01/01/2019.}}
   	\label{fig1}
   \end{figure}

In a possible  application, the observations are obtained from stock market returns. Concrete,  we consider  two stock price processes  and a time period $[0,T_0]$, where $T_0\in(0,\infty)$ is a time horizon. The stock price processes are denoted by
   \begin{equation*}
   \begin{split}
   Y_{i}=(Y_{i}(t);t\in [0,T_0]),~ i=1,2,
   \end{split}
   \end{equation*}
   where $Y_{i}$ is a random variable which takes values in the space of all measurable and square integrable  real-valued functions on $[0,T_0]$. Additional structural assumptions on the underlying stochastic process  for the stock prices are required. In the classical models for stock prices, i.e.,  the exponential L\'evy model,   Black-Scholes model, and  Merton model, the structural assumptions mentioned are independence and stationarity  of the increments of a  L\'evy process. Seasonality effects, that are specific trends during certain periods of the  stock price  processes, can disturb the stationarity  assumption. Figure \ref{fig1} shows the mean monthly indices (open) of the Nikkei Stock Average (Nikkei 225), Dow Jones Industrial Average (DJIA), and Standard \& Poor's 500 (S\&P 500)  from 01/01/1999 to 01/01/2019, where  seasonality effects are clearly seen. We will tackle this problem by splitting up the time horizon into $n$ equal sized time intervals to obtain $n$ observations for each stock price process. To be more specific, let $T_0=nT$ for $T\in(0,\infty)$ then we consider the time periods  $[0,T],\dots,[(n-1)T,nT]$ and our observations are the log-returns during these periods
   \begin{align*}
   		X_{j,i} (t)=\log  \frac{Y_{i}(t+(j-1)T)}{  Y_{i}((j-1)T)},~t\in[0,T],~i=1,2,~j=1,\dots,n,
   \end{align*} 
which are themselves measurable and square integrable real-valued functions on $[0,T]$. In particular, we have  the specific space $H=L^2[0,T]$ containing all measurable and square integrable real-valued functions on the interval $[0,T]$ of length $T\in(0,\infty)$ and  equipped with the usual inner product $\langle f, g \rangle = \int_0^T f(x)g(x) \,\mathrm{ d }x$, $f, g\in H$. A corresponding orthonormal basis is given by normalized Legendre polynomials. The  well-known models for stock prices  mentioned imply an independent and identical distributed structure of the increments $X_{j,i}$. In models with time-dependent or stochastic volatility (volatility clustering for instance), this structure is may be violated. In fact, the theory of $U$-Statistics is well-developed and  covers also  cases where the independent and identical distributed data structure is disturbed, see Chapters 2.3 and 2.4 of \cite{Lee1990}. We point out that our results are derived by application of the  theory of $U$-Statistics in the independent and identical distributed data case, but it should be possible to extend and modify the approach in more general situations, however under suitable regularity conditions.

\begin{figure}
   	\begin{minipage}{.5\textwidth}
   		\centering
   		\includegraphics[width=1.0\linewidth]{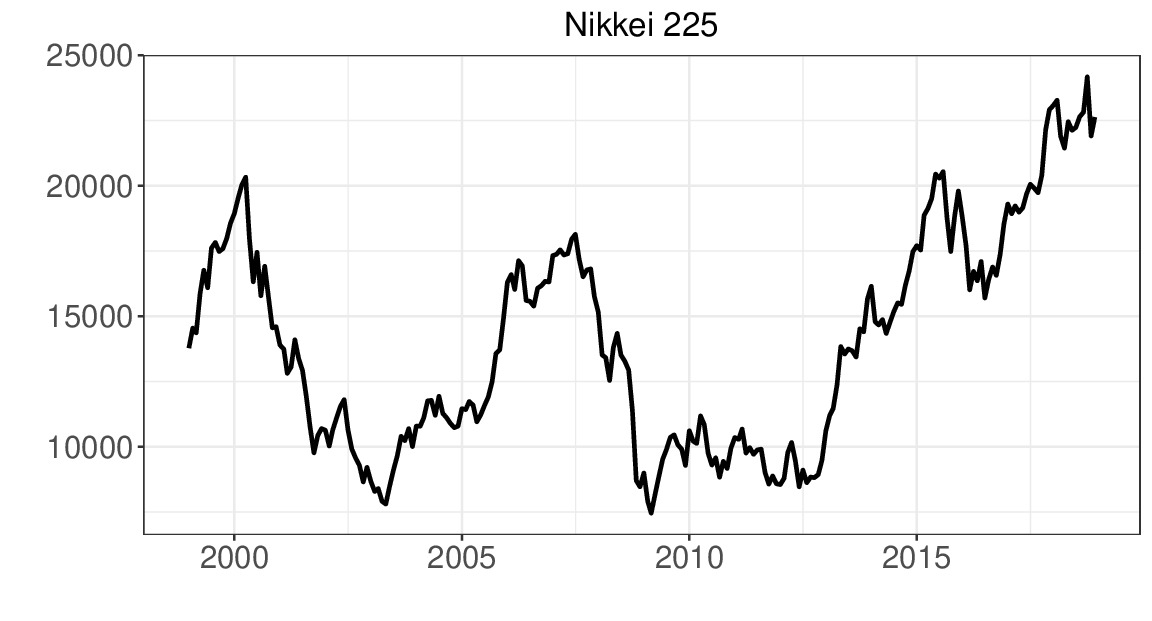}
   	\end{minipage}%
   	\begin{minipage}{.5\textwidth}
   		\centering
   		\includegraphics[width=1.0\linewidth]{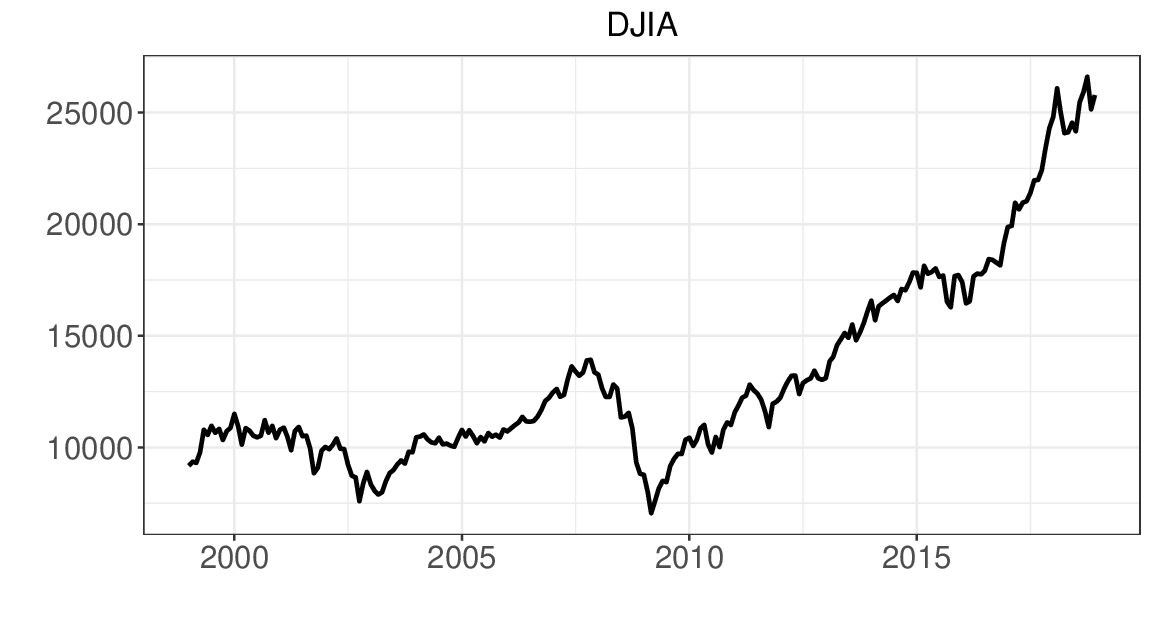}
   	\end{minipage}
   \centering
   	\begin{minipage}{.5\textwidth}
   		\centering
   		\includegraphics[width=1.0\linewidth]{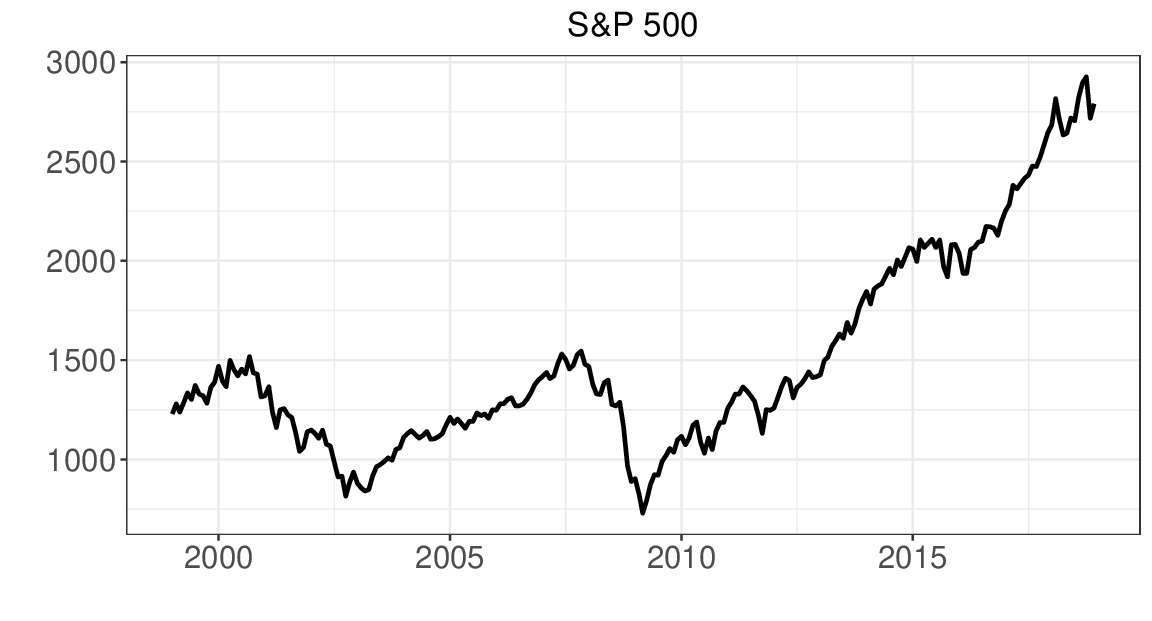}
   	\end{minipage}%
   	\caption{{Monthly values from 01/01/1999 to 01/01/2019.}}
   	\label{fig2}
   \end{figure}

\medskip
In what follows, we demonstrate the application of our test to the values (open) of the stock market indices Nikkei Stock Average of Japan, Dow Jones Industrial Average of US, and Standard \& Poor's 500 of US for the time period  01/01/1999 to 01/01/2019. For the demonstration how the test works in applications, we limit ourselves to the consideration of  the the monthly values  and a  linear interpolation. The resulting time series are presented in Figure \ref{fig2} and can be seen as square integrable functions on the interval $[0,T_0]$ for $T_0=20$ (years). To cover seasonality effects indicated by Figure \ref{fig1}, we split the time horizon of 20 years into 20 subintervals each representing one year, i.e. $T=1$ and $n=20$.  We apply our method to do pairwise comparisons of the indices, where the test statistic is again approximated by 500 random projects following {\textsc{Step}} 1--4 and the shifted Poisson distribution is used in {\textsc{Step}} 1 and {\textsc{Step}} 2 as in Section \ref{sec:sim}. The  resulting $p$-values for the bootstrap  approach are displayed in Table \ref{tab3} for 5000 resampling iterations, respectively. Since DJIA and S\&P 500 reflect both the US market, it is not surprising that the test leads to a very high $p$-value and, thus, do not reject the null hypothesis. Comparisons of each of these US indices with the Japanese Nikkei 225 lead to $p$-values around the typical used $5\%$-benchmark. This is inline with the first graphically impression, which we get by Figure \ref{fig2}.

	\renewcommand{\arraystretch}{1}
	\begin{table}[tb]
		\caption{\label{tab3}Empirical $p$-values of the test.}
		\begin{center}
			\begin{tabular}{@{}ccc@{}}
				$X_{1,j}(t)$&$X_{2,j}(t)$&   $p$-value  \\
				\hline
				DJIA & S\&P 500 &   73.6\\
				Nikkei 225 & S\&P 500 & 5.5\\
				Nikkei 225 & DJIA  &   5.6
			\end{tabular}
		\end{center}
	\end{table}


\begin{appendix}

\section{Proofs}

We  prove Theorem \ref{theo:asym_S} in a more general way. Instead of $\CvM_n$ we consider
 \begin{align*}
 	S_n= n\int  \int\left([  F_{1,n}(x,y)- F_1(x,y) ] - \left [  F_{2,n}(x,y) - F_2(x,y)\right]\right)^2\overline F_n(x,\mathrm d y)\mathcal P(\mathrm d   x),
 \end{align*}
where
\begin{equation*}
	\begin{split}
	F_{i }(x,y)=P({\langle x,X_{1,i}\rangle \le y}),~(x,y)\in H\times \R, ~i=1,2.
	\end{split}
	\end{equation*}
 
 \begin{theorem}\label{theo:asym_Sn}
 	Let $\tau_1,\tau_2,\ldots$ be a sequence of independent standard normal distributed random variables. Under the null hypothesis $\mathcal H$ as well as under any alternative, we have
 	\begin{align}\label{eqn:Sn_conv}
 	 S_n \overset{ d}{\rightarrow} \sum_{i=1}^\infty \lambda_i (1+\tau_i^2) = Z,
 	\end{align}
 	where $(\lambda_i)_{i\in\N}$ is a sequence of non-negative numbers with $\sum_{i=1}^\infty \lambda_i< \infty$ and $\lambda_{i}>0$ for at least one $i\in\N$ implying that  the distribution function of  $Z$ is continuous and strictly increasing on the non-negative half-line.
 \end{theorem}
 
 \textbf{Proof} Let $x_j=(x_{j,1},x_{j,2})\in H^2,~j\in \N$. We introduce the asymmetric kernel $f$ given by
 \begin{align}\label{eqn:def_unsym_kernel_f}
 &f(x_1,x_2,x_3) \\
 &= \frac{1}{2}\sum_{i=1}^2 \int \Bigl[ \I _{ \langle x_{1,1} - x_{3,i}, u \rangle \leq 0} - F_1(u, \langle x_{3,i}, u \rangle ) - \I _{ \langle x_{1,2} - x_{3,i}, u \rangle \leq 0  } + F_2(u, \langle x_{3,i}, u \rangle ) \Bigr]\nonumber \\
 &\times 
 \Bigl[ \I _{ \langle x_{2,1} - x_{3,i}, u \rangle \leq 0  } - F_1(u, \langle x_{3,i},u \rangle ) - \I _ { \langle x_{2,2} - x_{3,i}, u \rangle \leq 0 } + F_2(u, \langle x_{3,i}, u \rangle ) \Bigr] \mathcal P(\mathrm{ d }u) \nonumber 
 \end{align}
 as well as its symmetric version $\phi$ defined by
 \begin{align*}
 \phi(x_1,x_2,x_3) = \frac13 \Bigl( f(x_1,x_2,x_3)+f(x_2,x_3,x_1)+ f(x_1,x_3,x_2) \Bigr).
 \end{align*}
 Clearly,
 \begin{align*}
	 S_n&= \frac{1}{n^2}\sum_{i,  j,  k=1}^n f(X_i,X_{  j},X_{  k}) = \frac{1}{n^2}\sum_{i,  j,  k=1}^n \phi(X_i,X_{  j},X_{  k}).
 \end{align*}
 It is easy to check that for every $x_3\in H^2$ we have
 \begin{align}\label{eqn:deg_kernal_1RV}
 \E [f(X_1,x_2,x_3) ] = \E[ f(x_1,X_2,x_3) ] = 0.
 \end{align}
 The function $(x_1,x_2)\mapsto \E[\phi(x_1,x_2,X_3)]=\E [ f(x_1,x_2,X_3)] /3$ is not constant with probability one. This can easily been verified and also follows from our considerations below. Moreover, we can deduce from \eqref{eqn:deg_kernal_1RV} and the independence of the random variables that
 \begin{align}\label{eqn:deg_kernal_2RV}
 \E \big[ \phi(x_1,X_2,X_3) \big] = \E \big\{ \E\big[  \phi(x_1,X_2,X_3) \big | X_3\big] \big\} = 0 = \E \big[ \phi(X_1,X_2,X_3) \big].
 \end{align}
 In all, $\phi$ is degenerate of order $1$, see \cite{AcroneGine1992} for a detailed definition. Hence, we can deduce from Theorem 3.5 of \cite{AcroneGine1992} that $S_n-\E[S_n]$ converges in distribution to $Z$ as $n\to \infty$ if and only if 
 \begin{align*}
 V_n = 3n^{-1} \sum_{i,  j=1}^n \widetilde \phi (X_i,X_{  j})
 \end{align*}
 converges to $Z$ as well, where 
 \begin{align*}
 \widetilde \phi (x_1,x_2) &= \E \big[ \phi(x_1,x_2,X_3) \big] - \E \big[ \phi(x_1,X_2,X_3) \big] - \E \big[ \phi(X_1,x_2,X_3) \big] + \E \big[ \phi(X_1,X_2,X_3) \big].
 \end{align*}
 The new kernel $\widetilde \phi$ is a projection of $\phi$ to a corresponding function space, details are carried out in \cite{AcroneGine1992}. By \eqref{eqn:deg_kernal_1RV} and \eqref{eqn:deg_kernal_2RV} we can simplify it as follows
 \begin{align*}
 \widetilde \phi (x_1,x_2) = \E \big[ \phi(x_1,x_2,X_3) \big] = \frac{1}{3}\E \big[ f(x_1,x_2,X_3) \big].
 \end{align*}
 Thus, we can rewrite $ V_n$ as 
 \begin{align}\label{eqn:def_VN+tildef}
 	V_n = n^{-1} \sum_{i,  j=1}^n \widetilde f (X_i,X_{  j}) \text{ with }\widetilde f(x_1,x_2) = \E \big[ f(x_1,x_2,X_3) \big].
 \end{align}
 By Lemma \ref{lem:Mercer_kernel}, see below, $\widetilde f$ is a degenerated and bounded Mercer kernel. Due to the degeneracy of the kernel, the map $ g(\cdot)\longmapsto \E\big[\widetilde f(X_1,\cdot)g(X_1)\big]$  defines a Hilbert-Schmidt operator in the space of all square integrable functions on $H^2$ with respect to $P^{X_1}$, see also Section 4.3 in \cite{kor}. In this space, there exists an orthonormal basis of (centered) eigenfunctions $(\varphi_i)_{i\in\N}$ of this integral operator with corresponding non-negative eigenvalues $(\lambda_i)_{i\in\N}$. In detail, we have $\E[\varphi_i(X_1)]=0$, $\E[\varphi_i(X_1)\varphi_j(X_1)]=\I\{i=j\}$ and $\E[\varphi_i(X_1)\widetilde f(X_1,x)]= \lambda_i\varphi(x)$ for all $i,j\in\N$ and $x\in H^2$. Since $\widetilde f$ is a bounded Mercer kernel, we obtain, in analogy to the argumentation of \cite{leuchtneumann} in their proof of Theorem 2.1, from an extension of the Theorem of \cite{mercer} by \cite{sun} that for all $x,y$ in the support of  $P^{X_1}$
 \begin{align}\label{eqn:mercer}
 \widetilde f(x,y) = \sum_{i=1}^\infty \lambda_i\varphi_i(x)\varphi_i(y),
 \end{align}
 where the sum converges absolutely and uniformly on every compact subset of the cartesian square of the   support of  $P^{X_1}$. In particular, we obtain
 \begin{align}\label{nondegvor}
 \sum_{i=1}^\infty \lambda_i^2=\E[ \widetilde f(X_1,X_1)^2]<\infty~\text{and}~\sum_{i=1}^\infty \lambda_i=\E[	\widetilde f(X_1,X_1)]<\infty.
 \end{align}
 Regarding \eqref{eqn:mercer} we have
 \begin{align*}
 V_n= \sum_{i=1}^\infty \lambda_i\Bigl[ \frac{1}{\sqrt{n}}\sum_{j=1}^n\varphi_i(X_j)\Bigr]^2.
 \end{align*}
 From the orthogonality of $(\varphi_i)_{i\in\N}$ and the multivariate central limit theorem we can conclude that for each fixed $k\in\N$
 \begin{align*}
	\Bigl[ \frac{1}{\sqrt{n}}\sum_{j=1}^n\varphi_1(X_j),\ldots, \frac{1}{\sqrt{n}}\sum_{j=1}^n\varphi_k(X_j) \Bigr] \overset{ d}{\rightarrow}  (\tau_1,\ldots,\tau_k) \textrm{ as }n\to\infty,
 \end{align*}
 where $\tau_1,\tau_2,\ldots$ are independent and standard normal distributed. Combining this, \eqref{nondegvor} and a standard truncation argument, compare to Theorem 4.3.1 and 4.3.2 of \cite{kor} as well as the corresponding proofs, yields 
 \begin{align}\label{eqn:conv_Tn*}
 	 V_n \overset{ d}{\rightarrow}  \sum_{j=1}^\infty \lambda_j \tau_j^2  \textrm{ as }n\to\infty.
 \end{align}
 Now, note that by \eqref{eqn:deg_kernal_1RV}, \eqref{eqn:deg_kernal_2RV}, and \eqref{nondegvor}
 \begin{align*}
 &\E [S_n] \\
= &\frac{n}{n^2} \Big \{ \E[\phi(X_1,X_1,X_1)] + 3(n-1)\E[\phi(X_1,X_1,X_3)] +(n-1)(n-2)\E[\phi(X_1,X_2,X_3)] \Big\} \\
 = &\frac{1}{n}\E[\phi(X_1,X_1,X_1)] + \frac{3(n-1)}{n}\E[\phi(X_1,X_1,X_3)] \\
 &\rightarrow 3\E[\phi(X_1,X_1,X_3)] = \E[f(X_1,X_1,X_3)] =  \sum_{i=1}^\infty \lambda_i~\text{as}~n\to \infty.
 \end{align*}
 Consequently, \eqref{eqn:null_conv} follows. It  remains to show  that $\lambda_i>0$ for at least one $i\in\N$.  Let us suppose  that $\lambda_i=0$ for all $i\in\N$ for a moment. Then, it follows from \eqref{eqn:null_conv} that $\CvM_n\overset{ P}{\rightarrow}  0$ as $n\to\infty$. Moreover, our assumptions on  the joint distribution of $\langle X_{1,1},e_i\rangle$ and $\langle X_{1,2},e_i\rangle$, $i\in I$, ensure that Assumption 1 and Assumption 2 in \cite{gai} are satisfied, and it follows from Theorem 1 and Theorem 3 in \cite{gai} that $D_n(e_1) \overset{ d}{\rightarrow} S$ as $n\to\infty$, where $S$ is a real-valued random variable and not constantly zero. In all, from
	\begin{equation*}
		\CvM_n\ge  D_n(e_1)\nu_1(\{1\})  \overset{ d}{\rightarrow}  S\cdot\nu_1(\{1\})  \textrm{ as }n\to\infty
	\end{equation*}
together with $\nu_1(\{1\})>0$ we obtain a contradiction. This completes the proof. \hfill $\square$
 \\
 
  \begin{lemma}\label{lem:Mercer_kernel}
 	The function $\widetilde f$ is a degenerated and bounded Mercer kernel, i.e., it is continuous, symmetric and positive semidefinite.
 \end{lemma}
 \textbf{Proof}
 It is easy to see that $\widetilde f$ is bounded and symmetric. The degeneracy follows immediately from \eqref{eqn:deg_kernal_1RV}. For arbitrary $k\in\N$ let $c_1,\ldots,c_k\in\R$. Then
  \begin{align*}
	 \sum_{i,j=1}^k c_ic_j \widetilde f(x_i,x_j) 
	 &= \frac{1}{2}\sum_{\ell=1}^2 \E \Bigl[ \int \Bigl\{ \sum_{i=1}^k c_i \Bigl[ \I _{ \langle x_{i,1}, x \rangle \leq \langle X_{3,\ell}, x \rangle } - F_1\big(x, \langle X_{3,\ell}, x \rangle \big)  \\
	 &\phantom{=}-\I _{\langle x_{i,2}, x \rangle \leq \langle X_{3,\ell}, x \rangle} + F_2(x, \langle X_{3,\ell}, x \rangle ) \Big] \Bigr\}^2 \mathcal P(\mathrm{ d }x) \Bigr] \geq 0.
 \end{align*}
 Hence, $\widetilde f$ is positive semidefinite. For the continuity proof, let $(x_{1n})_{n\in\N}$ and $(x_{2n})_{n\in\N}$ be sequences in $ H^2$ such that  $\lim_{n\to\infty}x_{jn}=x_j\in H^2$, $j=1,2$. By Lemma 3.1 of \cite{DitzhausGaigall2018} 
 \begin{align}\label{eqn:ditz+gaig_intI=0}
 \I_{ \langle x, x_{3,\ell} - y \rangle \neq 0 } = 1 
 \end{align}
 for $\mathcal P \times P^{X_{1,\ell}}$-almost all $(x,x_{3,\ell})$ and every $y \in  H$. This and the continuity of the inner product imply for $\mathcal P$-almost all $x$ and every $m,\ell \in\{1,2\}$ that
 \begin{align*}
 \lim_{n\to\infty}\I_{ \langle x, x_{jn,m}\rangle \leq \langle x, X_{3,\ell}  \rangle } = \I_{\langle x, x_{j,m} \rangle \leq \langle x, X_{3,\ell}  \rangle }\text{ with probability one}.
 \end{align*}
 Consequently, $\widetilde f(x_{1n},x_{2n})$ converges to $\widetilde f(x_{1},x_{2})$. \hfill $\square$  \\

\textbf{Proof of Theorem \ref{theo:asym_S}} Since $F_1=F_2$ and, thus, $S_n=\CvM_n$ under the null hypotheses, the statement follows immediately from Theorem \ref{theo:asym_Sn}. \hfill $\square$ \\
 
\textbf{Proof of Theorem \ref{theo:cons_S}} First, observe that
 \begin{align*}
 &n^{-1}\CvM_n = n^{-1}S_n  \\
 &+2  \int  \int[  F_{n,1}(x,y)- F_1(x,y) -  F_{n,2}(x,y) + F_2(x,y)][  F_1(x,y)  - F_2(x,y)]\bar F_n(x,\mathrm d y)\mathcal P(\mathrm d   x)\\
 & + \int  \int[ F_1(x,y) -  F_2(x,y)]^2\bar F_n(x,\mathrm d y)\mathcal P(\mathrm d   x).
 \end{align*}
 By Theorem \ref{theo:asym_Sn}, $n^{-1}S_n$ converges in probability to $0$.  By the Cauchy-Schwarz inequality the absolute value of the second summand is bounded from above by $2\sqrt{n^{-1}S_n}$. In particular, the second summand vanishes in probability as well. The third summand can be rewritten as
 \begin{align*}
 \frac{1}{2n}\sum_{j=1}^n \sum_{k=1}^2 \int \Big[ F_1(x,\langle x, X_{j,k}\rangle) -  F_2(x,\langle x, X_{j,k}\rangle)\Big]^2\mathcal P(\mathrm d   x) = \frac{1}{2n}\sum_{j=1}^n g(X_j),
 \end{align*}
 for an appropriate function $g$. By the strong law this sum converges almost surely to
 \begin{align}
 &\frac{1}{2}\sum_{k=1}^2 \int  \E \Bigl[ \Big\{ F_1(x,\langle x, X_{3,k}\rangle) -  F_2(x,\langle x, X_{3,k}\rangle)\Big\}^2 \Bigr]\mathcal P(\mathrm d   x)\nonumber \\
 &= \frac{1}{2}\sum_{k=1}^2 \int \big[ F_1(x,y) -  F_2(x,y)\big]^2 Q_k[\mathrm d   (x,y)]\label{eqn:cons_simpl_3sum},
 \end{align}
 where $Q_k$ is the distribution introduced at the proof's end of Theorem \ref{theo:asym_Sn}. In analogy to the argumentation of \cite{DitzhausGaigall2018} in the proof for their Theorem 3.2, we can conclude that each summand from \eqref{eqn:cons_simpl_3sum} is strictly positive. Finally, we obtain
 \begin{align*}
	 	n^{-1}\CvM_n \overset{p}{\rightarrow}  \frac{1}{2}\sum_{k=1}^2 \int \big[ F_1(x,y) -  F_2(x,y)\big]^2 Q_k(\mathrm d (x, y)) > 0.
 \end{align*}
{}\hfill $\square$\\

 \textbf{Proof of Theorem \ref{theo:boot}} From now on, we suppose that the data $X_1,\ldots,X_n$ are fixed and we operate on the conditional space. In particular, we can treat $F_{n,i}$, $i=1,2$, as a non-random function, which converges, without loss of generality, pointwisely to $F_i$. First, we remark that the distribution of the bootstrap sample depends on the sample size. Moreover, the distribution of $X_{in}^*$ converges weakly to the distribution of $X_i$. Thus, by Theorem 1.10.4 of \cite{van} we can assume without loss of generality that $X_{in}^*$ converges to $X_i'$ for all $i\in \N$ with probability one, where $X_i'$ has the same distribution as $X_i$, and that $X_r'$ is independent from $X_{1n}^*,\ldots,X_{(r-1)n}^*,X_{(r+1)n}^*,\ldots$ for all $r\in\N$. Now, define
 \begin{align*}
 &f_{n}^*(x_{1},x_{2},x_{3}) \\
 &= \frac{1}{2}\sum_{k=1}^2 \int \Bigl[ \I_{ \langle x, x_{1,1} -x_{3,k} \rangle \leq 0 } -  F_{n,1}(x, \langle x,x_{3,k}\rangle ) - \I _{\langle x,x_{1,2} - x_{3,k} \rangle \leq  0} +  F_{n,2}(x, \langle x, x_{3,k} \rangle ) \Bigr]\\
 &\times 
 \Bigl[ \I _{ \langle x,x_{2,1} - x_{3,k}\rangle \leq 0  } -  F_{n,1}(x, \langle x, x_{3,k} \rangle ) - \I_{ \langle x, x_{2,2} - x_{3,k} \rangle \leq 0 } +  F_{n,2}(x, \langle x, x_{3,k}\rangle ) \Bigr] \mathcal P(\mathrm{ d }x).
 \end{align*}
 Then we have
 \begin{align*}
 \CvM_{n}^* = \frac{1}{n^2} \sum_{i,j,k=1}^n f_{n}^*(X_{in}^*,X_{jn}^*,X_{kn}^*).
 \end{align*}
 Define
 \begin{align*}
 S_n' = \frac{1}{n^2} \sum_{i,j,k=1}^n f( X_{i}',  X_{j}', X_{k}')~\text{and}~\kappa_{n,i,j,k} = f_{n}^*(X_{in}^*,X_{jn}^*,X_{kn}^*) - f( X_i',  X_j',  X_k'),
 \end{align*}
 where $f$ is defined in \eqref{eqn:def_unsym_kernel_f}. By Theorem \ref{theo:asym_Sn}, $ S_n'$ converges in distribution to $Z$.  Combining this and 
 \begin{align}\label{eqn:boot_null_suff}
 \E[ (\CvM_{n}^* -  S_n')^2] &= n^{-4}\sum_{i_1,\ldots,i_6=1}^n \E[\kappa_{n,i_1,i_2,i_3}\kappa_{n,i_4,i_5,i_6}] \rightarrow 0~\text{as}~n\to \infty
 \end{align}
 under the null hypothesis, where the proof of \eqref{eqn:boot_null_suff} is given later, yields conditional convergence
 \begin{align*}
 	\CvM_{n}^* \overset{ d}{\rightarrow} Z \textrm{ as }n\to\infty
	\end{align*}
	given the observations $X_1,\ldots,X_n$ under $\mathcal H$ for $Z$ from Theorem \ref{theo:asym_S}. Consequently, we can deduce that $c_{n,1-\alpha}^* \overset{p}{\rightarrow} c_{1-\alpha}$ under $\mathcal H$ and, in particular, the statement under $\mathcal H$ follows, compare to Lemma 1 of \cite{janssenpauls}. 
	For the statement under the alternative it remains to show that $(\CvM_{n}^*)_{n\in\N}$ is a tight sequence of real valued random variables, compare to Theorem 7 of \cite{janssenpauls}, i.e. we have to show 
	\begin{align*}
		\limsup_{K\to\infty}\limsup_{n\to\infty} P( |\CvM_{n}^*|\geq K) =0.
	\end{align*}
	 In contrast to \eqref{eqn:boot_null_suff}, it remains now to show 
	 \begin{align}\label{eqn:boot_alt_suff}
	 	\limsup_{n\to\infty}\E[ ( \CvM_{n}^* -  S_n')^2] \leq M < \infty.
	 \end{align}
	To sum up, we need to verify \eqref{eqn:boot_null_suff} for the statement under $\mathcal H$ and \eqref{eqn:boot_alt_suff} for the statement under $\mathcal K$. For this purpose, we divide the corresponding sum in \eqref{eqn:boot_null_suff} into the following six sums
 \begin{align*}
 	I_{n,p} = n^{-4}\sum_{i_1,\ldots,i_6=1}^n \E[\kappa_{n,i_1,i_2,i_3}\kappa_{n,i_4,i_5,i_6}] \I_{ |\{i_1,\ldots,i_6\}| = p },~p=1,\ldots,6.
 \end{align*}
 First, we will prove that $I_{n,p}$ converges to $0$ for $p\in\{1,2,3,5,6\}$ independently whether the null hypothesis or the alternative is true. At the end, we discuss $I_{n,4}$ separately under the null hypothesis and the alternative.  For all considerations below, remind that $\kappa_{n,i,j,m}$ is uniformly bounded by $8$. As a first consequence of this, we obtain that as $n\to \infty$
 \begin{align*}
 I_{n,1} + I_{2,n} + I_{3,n} \leq \frac{8^2}{n^4}[ n + (2^6-1)n(n-1)+(3^6-2^6)n(n-1)(n-2) ]\rightarrow 0.
 \end{align*}
 Let us have now a look on all summands with $|\{i_1,\ldots,i_6\}|=5$. Let $r$ be the number that appears twice within the indices $i_1,\ldots,i_6$. Observe that \eqref{eqn:deg_kernal_1RV} also holds for the bootstrap sample, i.e.,
 \begin{align*}
 \E [ f_{n}^*(X_{1n}^*,x_{2},x_{3}) ] = \E[ f_{n}^*(x_{1},X_{2n}^*,x_{3}) ] = 0.
 \end{align*}
 Combining this and \eqref{eqn:deg_kernal_1RV} yields $ \E [ \kappa_{n,i,j,k}  |  X_r',X_{r}^* ] = 0$ with probability one whenever $|\{i,j,k\}|=3$. Consequently,
 \begin{align*} 
 \E[ \kappa_{n,i_1,i_2,i_3}\kappa_{n,i_4,i_5,i_6} ] = \E\{ \E [ \kappa_{n,i_1,i_2,i_3}  |  X_r',X_{r,*} ]\E [ \kappa_{n,i_4,i_5,i_6} |  X_r',X_{r,*}] \} = 0
 \end{align*}
 Clearly, the same can be shown for the case  $|\{i_1,\ldots,i_6\}|=6$. Hence, $I_{n,5}+I_{n,6}=0$.
 
 Now, we consider $I_{n,4}$. Due to the boundedness of $\kappa_{n,i,j,m}$, we always obtain
 \begin{align*}
 I_{n,4} \leq \frac{8^2}{n^4}(4^6-3^6)n(n-1)(n-2)(n-3) \leq 2^{20}.
 \end{align*}
 From this and the previous considerations we can conclude \eqref{eqn:boot_alt_suff}. Now, let us suppose that the null hypothesis  is true. Due to symmetry $\kappa_{n,i,j,k}=\kappa_{n,j,i,k}$ we get 
 \begin{align*}
 I_{n,4} \leq \frac{8}{n^4}(4^6-3^6)n(n-1)(n-2)(n-3) \max\Bigl\{ \E[|\kappa_{n,1,1,2}|]+\E[|\kappa_{n,1,2,2}|] + \E[|\kappa_{n,1,3,2}|] \Bigr\}.
 \end{align*}
 Consequently, it is sufficient for \eqref{eqn:boot_null_suff} to prove
 \begin{align}\label{eqn:boot_I4_suff}
 \lim_{n\to\infty}\kappa_{n,1,j,2} = 0 ~ \text{in probability}
 \end{align}
 for $j=1,2,3$. From the continuity of the inner product, \eqref{eqn:ditz+gaig_intI=0}, the underlying independence and the convergence of $X_{1n}^*$, $X_{2n}^*$, $X_{3n}^*$ we obtain that with probability one
 \begin{align}\label{eqn:boot_intI=0}
 \lim_{n\to\infty}\I_{ \langle x,  X_{rn,k}^*  \rangle \leq \langle x, X_{2n,l}^* \rangle } =  \I_{\langle x,  X_{r,k}' \rangle \leq \langle x,  X_{2,l}'\rangle } \text{ for }\mathcal P\text{-almost all }x,
 \end{align}
 every $r\in\{1,3\}$ and $k,l\in\{1,2\}$. Analogously, we have 
 \begin{align}\label{eqn:boot_intF=0}
 	&F_{n,k}( x,\langle x,  X_{2n,l}^* \rangle ) =  \E\Bigl[\I_{ \langle x,  X_{1n,k}^*  \rangle \leq \langle x,  X_{2n,l}^*\rangle }  | X_{2n,l}^* \Bigr] \nonumber \\
 	&\overset{a.s.}{\rightarrow} \E\Bigl[ \I_{ \langle x,  X_{1,k}'  \rangle \leq \langle x,  X_{2,l}' \rangle }  | X_{2,l}' \Bigr] = F_k( x,\langle x,  X_{2,l}' \rangle )~\text{as}~n\to \infty.
 \end{align}
 Combining \eqref{eqn:boot_intI=0} and \eqref{eqn:boot_intF=0} shows \eqref{eqn:boot_I4_suff} for the case $j\in\{1,3\}$. The reason why we need to be more careful in the case $j=2$ is that, in general, \eqref{eqn:boot_intI=0} is false if $r=j=2$. However, the integrals appearing in the limiting $f( X_1', X_2', X_2')$ vanish when they are restricted to the crucial (random) set $A=\{x:\langle x,  X_{2,2}' -  X_{2,1}'\rangle = 0\}$. To be more specific, since the null hypothesis is true and, hence, $F_1=F_2$, we obtain 
 \begin{align*}
 &\sum_{k=1}^2 \int_A \Bigl[ \I  _{\langle x,  X_{1,1}' - X_{2,k}'\rangle \leq 0  } - F_1(x, \langle x, X_{2,k}'\rangle ) - \I _{ \langle x, X_{2,2}' - X_{2,k}'\rangle \leq 0 }+ F_2(x, \langle x,  X_{2,k}' \rangle ) \Bigr]\\
 &\times 
 \Bigl[ \I _{\langle x, X_{2,1}' -  X_{2,k}'\rangle \leq 0  } - F_1(x, \langle x,  X_{2,k}' \rangle ) - \I _{\langle x,  X_{2,2}' -  X_{2,k}' \rangle \leq 0 } + F_2(x, \langle x,  X_{2,k}' \rangle ) \Bigr] \mathcal P(\mathrm{ d }x)\\
 &=2\int_A \Bigl[ \I_{\langle x,  X_{1,1}' - X_{2,1}' \rangle \leq 0 }- F_1(x, \langle x, X_{2,1}' \rangle ) - \I_{ \langle x, X_{1,2}' - X_{2,1}' \rangle \leq 0 } + F_1(x, \langle x,  X_{2,1}' \rangle ) \Bigr]\\
 &\times 
 \Bigl[ F_1(x, \langle x,  X_{2,1}' \rangle ) - F_1(x, \langle x,  X_{2,1}' \rangle )   \Bigr] \mathcal P(\mathrm{ d }x) =0.
 \end{align*}
 Thus, \eqref{eqn:boot_I4_suff} follows again from \eqref{eqn:boot_intI=0}, \eqref{eqn:boot_intF=0}, the continuity of the inner product and the convergence of $X_{1n}^*$ and $X_{2n}^*$. {}\hfill $\square$\\

\end{appendix}


\begin{thebibliography}{}
    	
    	
    	\bibitem[Arcones and Gin\'e(1992)]{AcroneGine1992}
    	Arcones, M. A. and E. Gin\'e (1992). 
    	\newblock On the bootstrap of U and V statistics. 
    	\newblock {\em Annals of Statistics 20}, 655--674.


    	\bibitem[Ahmed(2017)]{ahmed2017}
    	Ahmed, S.E. (2017).
    	\newblock {\em Big and Complex Data Analysis. Contributions to Statistics}.
    	\newblock Cham: Springer.
    	
  
  	\bibitem[Anderson and Darling(1952)]{and}
    	{Anderson, T. W. and D. A. Darling} (1952).
    	\newblock Asymptotic Theory of Certain "Goodness of Fit" Criteria Based on Stochastic Processes. 
    	\newblock {\em Annals of Mathematical Statistics 23}, 193--212. 
    	

    	
    	
    	\bibitem[Bugni and Horowitz(2018)]{BugniHorwitz2018}
    	Bugni, F. A. and J. L. Horowitz (2018). 
    	\newblock Permutation Tests for Equality of Distributions of Functional Data. 
    	\newblock {\em ArXiv e-prints} (arXiv:1803.00798).
    	
    	\bibitem[Bugni et al.(2009)]{bun}
    	Bugni, F. A., P. Hall, J. L. Horowitz and G. R. Neumann (2009).
    	\newblock Goodness-of-fit tests for functional data. 
    	\newblock{\em Econometrics Journal  12}, S1--S18. 
    	
    	\bibitem[Cuesta-Albertos et al.(2006)]{cue2006}
    	{Cuesta-Albertos, J. A. and R. Fraiman and T. Ransford} (2006). 
    	\newblock Random projections and goodness-of-fit tests in infinite-dimensional spaces. 
    	\newblock {\em Bull. Braz. Math. Soc. (N.S.) 37}, 477--501.
    	
    	
    	\bibitem[Cuesta-Albertos et al.(2007)]{cue2007}
    	{Cuesta-Albertos, J. A., E. del Barrio, R. Fraiman, R. and C. Matr\'an} (2007). 
    	\newblock The random projection method in goodness of fit for functional data. 
    	\newblock{\em Computational Statistics \& Data Analysis 51}, 4814--4831.
    	
    	
    	\bibitem[Cuesta-Albertos and Febrero-Bande(2010)]{cue2010}
    	{Cuesta-Albertos, J. A. and M. Febrero-Bande} (2010). 
    	\newblock A simple multiway ANOVA for functional data. 
    	\newblock {\em TEST 19}, 537--557. 
    	
    	
    	\bibitem[Cuevas and Fraiman(2009)]{cuevas}
    	{Cuevas, A. and R. Fraiman} (2009). 
    	\newblock On depth measures and dual statistics. A methodology for dealing with general data. 
    	\newblock {\em Journal of Multivariate Analysis 100}, 753--766. 
    	
    	
    	\bibitem[Cuevas(2014)]{cuevas14}
    	{Cuevas, A.} (2014). 
    	\newblock A partial overview of the theory of statistics with functional data. 
    	\newblock {\em Journal of Statistical Planning and Inference 147}, 1--23. 
    	
    	
    	\bibitem[Ditzhaus and Gaigall(2018)]{DitzhausGaigall2018}
    	Ditzhaus, M. and D. Gaigall (2018).
    	\newblock A consistent goodness-of-fit test for huge dimensional and functional data.
    	\newblock {\it Journal of Nonparametric Statistics 30}, 834--859.
    	
    	
    	\bibitem[Efron(1979)]{efron}
    	{Efron, B.} (1979).
    	\newblock Bootstrap methods: Another look at the jackknife.
    	\newblock {\em Annals of Statistics 7}, 1--26. 
    	

    	
    			


    	\bibitem[Gaigall(2019)]{gai}
    	 Gaigall, D. (2019)
    	\newblock Testing marginal homogeneity of a continuous bivariate distribution with possibly incomplete paired data. 
    	\newblock {\em Metrika 83}, 437--465.



    	\bibitem[Gray and French(2008)]{gra}
    Gray, B.J.,  French, D.W.  (2008)
    	\newblock Empirical comparisons of distributional models for stock index returns. 
    	\newblock {\em Journal of Business Finance \& Accounting 17}, 451--459.


    	\bibitem[G\"onc\"u et al.(2016)]{goe}
    	{G\"onc\"u, A.,  Oguz, M., Karahan, M.O., Kuzubas, T.U.} (2016). 
    	\newblock A comparative goodness-of-fit analysis of distributions of some Lévy processes and Heston model to stock index returns. 
    	\newblock {\em The North American Journal of Economics and Finance 36}, 69--83.
    	


    	
    	\bibitem[González-Manteiga and Crujeiras(2013)]{gon}
    	{González-Manteiga, W. and R. M. Crujeiras} (2013). 
    	\newblock An updated review of Goodness-of-Fit tests for regression models. 
    	\newblock {\em TEST 22}, 361--411.
    	



    	\bibitem[Goia and Vieu(2016)]{goiaVieu2016}
    	Goia, A., and P. Vieu (2016)
    	\newblock An Introduction to Recent Advances in High/Infinite Dimensional Statistics. 
    	\newblock {\em Journal of Multivariate Analysis 146}, 1–6.

    	    
        	\bibitem[Hall and Van Keilegom(2002)]{hal2017}
	    {Hall, P. and I. Van Keilegom} (2007).  
	    \newblock Two-sample tests in functional data analysis starting from discrete data.
	    \newblock{\em Statistica Sinica 17}, 1511--1531. 
    	
    	
    	\bibitem[Hall and Tajvidi(2002)]{hal}
    	{Hall, P. and N. Tajvidi} (2002).  
    	\newblock Permutation Tests for Equality of Distributions in High-Dimensional Settings. 
    	\newblock{\em Biometrika 89}, 359--374. 
    	

    	
    	    	
    	\bibitem[Janssen and Pauls(2003)]{janssenpauls}
    	Janssen, A. and T. Pauls (2003).
    	\newblock How do bootstrap and permutation tests work?
    	\newblock \emph{Annals of Statistics 31}, 768--806.
    	
    	\bibitem[Jiang, Meintanis and Zhu(2017)]{jiang2017}
    	Jiang, Q., Meintanis, S. G. and Zhu, L. (2017).
    	\newblock Two-sample tests for multivariate functional data.
    	\newblock In:  Aneiros G., Bongiorno E., Cao R., Vieu P. (eds) Func-
    	tional Statistics and Related Fields. Contributions to Statistics. Springer.
    	
    	\bibitem[Koroljuk and Borovskich(1994)]{kor}
    	{Koroljuk, V. S. and X. V. Borovskich, X.V.} (1994). 
    	\newblock {\em Theory of $U$-Statistics}.  
    	\newblock Dordrecht: Kluwer Academic Publishers Group.
    	
    	
    	\bibitem[Lee(1990)]{Lee1990}
    	{Lee, A. J.} (1990). 
    	\newblock {\em U-statistics: Theory and Practice}.  
    	\newblock Boca Ration, FL: CRC Press.
    	
    	
    	\bibitem[Leucht and Neumann(2013)]{leuchtneumann}
    	{Leucht, A. and M. H. Neumann} (2013). 
    	\newblock Dependent wild bootstrap for degenerate $U$- and $V$-statistics. 
    	\newblock{\em Journal of Multivariate Analysis 117}, 257--280.
    	
    	

    	\bibitem[Malevergne, Pisarenko, and Sornette(2005)]{mal}
    	Malevergne, Y.,  Pisarenko, V., Sornette, D. (2005)
    	\newblock Empirical distributions of stock returns: between the stretched exponential and the power law? 
    	\newblock {\em Quantitative Finance 5}, 379--401.


    	\bibitem[Mercer(1909)]{mercer}
    	{Mercer, J.} (1909). 
    	\newblock Functions of positive and negative type and their connection with the theory of integral equations.
    	\newblock {\it Philosophical Transactions of the Royal Society A 209}, 415--446.
    	
	\bibitem[Midesia et al.(2016)]{mid}
    	{Midesia, S., Basri, H.,  Shabri, M. Majid, M.S.A.} (2016). 
    	\newblock The Effects of Asset Management and Profitability on Stock Returns: A Comparative Study between Conventional and Islamic Stock Markets in Indonesia.
    	\newblock {\it Academic journal of economic studies 2}, 44--54.
    	
    	\bibitem[Min(2015)]{min}
    	{Min, S.} (2015). 
    	\newblock Goodness of Fit and Independence Tests for Major 8 Companies of Korean Stock Market.
    	\newblock {\it Korean Journal of Applied Statistics 28}, 1245--1255.




    	\bibitem[Rosenblatt(1952)]{ros}
    			{Rosenblatt, M.} (1952). 
    			\newblock Limit theorems associated with variants of the von Mises statistic. 
    			\newblock {\em Annals of Mathematical Statistics 23}, 617-623.    			
    			
    	\bibitem[Serfling(2001)]{serf}
    	{Serfling, R.S.} (2001). 
    	\newblock {\em Approximation Theorems of Mathematical Statistics}. 
    	\newblock New York, NY: Wiley. 
    			
    	\bibitem[Sun(2005)]{sun}
    			{Sun, H.} (2005). 
    			\newblock Mercer theorem for RKHS on noncompact sets. 
    			\newblock {\it Journal of Complexity 21}, 337--349.
    			
    	\bibitem[Van der Vaart and Wellner(1996)]{van}
    		{Van der Vaart, A. and J. A. Wellner} (1996).
    			\newblock {\em Weak convergence and empirical processes. With applications to statistics}. 
    			\newblock New York, NY: Springer.


    \end{thebibliography}
\end{document}